\documentclass[vecphys]{svmult}

\usepackage{makeidx}
\usepackage{graphicx}
\usepackage{multicol}
\usepackage[bottom]{footmisc}

\usepackage[utf8]{inputenc}
\usepackage{amsmath}

\makeindex

\begin{document}

\title*{Emergent Community Structure\\in Social Tagging Systems}

\author{Ciro Cattuto\inst{1,2,3}\and Andrea Baldassarri\inst{2}\and \\
Vito D.P. Servedio\inst{2,1}\and Vittorio Loreto\inst{2,3} }

\institute{Museo Storico della Fisica e Centro Studi e Ricerche ``Enrico Fermi'' \\
Compendio Viminale, 00184 Roma, Italy\\
\texttt{ciro.cattuto@roma1.infn.it}
\and Dipartimento di Fisica, Universit\`{a} di Roma ``La Sapienza'' \\
P.le A. Moro, 2, 00185 Roma, Italy \and
Institute for Scientific Interchange (ISI), Torino, Italy}

%\date{\today}

\maketitle

\begin{abstract}
A distributed classification paradigm known as \textit{collaborative
tagging} has been widely adopted in new web
applications designed to manage and share online resources.
Users of these applications organize resources
(web pages, digital photographs, academic papers)
by associating with them freely chosen text labels, or \textit{tags}.
Here we leverage the social aspects of collaborative tagging
and introduce a notion of \textit{resource distance}
based on the collective tagging activity of users.
We collect data from a popular system and perform experiments
showing that our definition of distance can be used to build
a weighted network of resources with a detectable
community structure. We show that this community structure
clearly exposes the semantic relations among resources.
The communities of resources that we observe are a genuinely
emergent feature, resulting from the uncoordinated activity
of a large number of users, and their detection paves the way
to mapping emergent semantics in social tagging systems.
\end{abstract}

\section{Introduction}
\label{sec:intro}
Information systems on the World Wide Web have been increasing in size
and complexity to the point that they presently exhibit features
typically attributed to \textit{bona fide} complex systems. They
display rich high-level behaviors that are causally connected in
non-trivial ways to the dynamics of their interacting elementary parts. 
Because of this, concepts and formal tools from the science
of complex systems can play an important role in understanding
the structure and dynamics of such systems.  

This study focuses on the recently established paradigm of
collaborative tagging~\cite{mates,connotea1}.
In web applications like \textit{del.icio.us}\footnote{\tt
http://del.icio.us/}, \textit{Flickr}\footnote{\tt
http://flickr.com/}, \textit{BibSonomy}\footnote{\tt
http://www.bibsonomy.org/} users organize diverse \textit{resources}
-- ranging from web pages to academic papers and photographs --
with semantically meaningful information in the form of text labels,
or ``tags''. Tags are freely chosen and users associate resources
with them in a totally uncoordinated fashion. 
Nevertheless, the tagging activity of each user is globally
visible to the user community and the tagging process develops
genuine social aspects and complex interactions
\cite{huberman,noi_tagging},
eventually leading to a bottom-up categorization of resources
shared throughout the user community.
The open-ended set of tags used within the system --
commonly referred to as ``folksonomy'' -- can be used as a sort of
semantic map to navigate the contents of the system itself.

\begin{figure}[htb]
\begin{center}
\includegraphics[width=0.95\columnwidth]{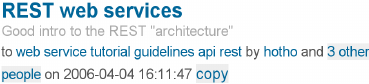}
\end{center}
\caption{The basic unit of information in a folksonomy, i.e. a post,
  is shown as it appears in the interface of {\tt bibsonomy.org} a
  social collaborative tagging system for bookmarks and scientific
  references. At the top, the title of the resource (a web page) is
  shown, followed by its own subtititle. Then the list of tags
  associated by the user {\tt hotho} is displayed. Other informations
  are: the number of other users who inserted the same resource in the
  system, as well as the date and time of insertion of the present
  post.
\label{bibsonomy_detail}}
\end{figure}

In figure~\ref{bibsonomy_detail} a single annotation
example (said ``post'') is shown, as appears in the interface of {\tt
  bibsonomy.org} system.

Our work is based on experimental data from one of the largest
and most popular collaborative tagging systems, \textit{del.icio.us},
currently used by over a million users to manage and share
their collections of web bookmarks.

%%%%%%%%%%%%%%%%%%%%%%%%%%%%%%%%%%%%%%%%%%%%%%%%%%%%%%%%%
%%%%%%%%%%%%%%%%%%%%%%%%%%%%%%%%%%%%%%%%%%%%%%%%%%%%%%%% 
%%%%%%%%%%%%%%%%%%%%%%%%%%%%%%%%%%%%%%%%%%%%%%%%%%%%%%%%%

The main point of our work is neither to present a new spectral community
detection algorithm, nor to report a large data set analysis.
Rather, we want to show that, choosing the right projection and the
right weighting procedure, we can produce a weighted undirected
network of resources from the full tri-partite folksonomy network, 
which embed a meaningful social classification of resources.
This is especially surprising, considering that users annotate
resources in a very anarchic, uncoordinated and noisy way.
%%%%%%%%%%%%%%%%%%%%%%%%%%%%%%%%%%%%%%%%%%%%%%%%%%%%%%%%%
%%%%%%%%%%%%%%%%%%%%%%%%%%%%%%%%%%%%%%%%%%%%%%%%%%%%%%%% 
%%%%%%%%%%%%%%%%%%%%%%%%%%%%%%%%%%%%%%%%%%%%%%%%%%%%%%%%%

In section~\ref{sec:data} we describe the experimental data
we collected. In Section~\ref{sec:distance} we introduce a notion
of resource distance based on the collective activity of users.
Based on that, we set up an experiment using actual data
from \textit{del.icio.us} and we build a weighted network of resources.
In section~\ref{sec:communities} we show that spectral methods
from complex networks theory can be used to detect clusters of resources
in the above network and we characterize those clusters
in terms of user tags, exposing semantics.
Finally, section~\ref{sec:conclusions} gives an overview of our results
and points to directions for future work.

\section{Experimental Data}
\label{sec:data}
Our analysis focuses on \textit{del.icio.us} for several reasons:
i) it was the first system to deploy the ideas of collaborative tagging
on a large scale, so it has acquired a paradigmatic character
and it is the natural starting point for any quantitative study.
ii) it has a large user community and contains a huge amount
of raw data on the structure and dynamics of a folksonomy.
iii) it is a \textit{broad folksonomy}~\cite{vanderwal},
i.e. single tag associations by different users retain their identity
and can be individually retrieved. This allows us to measure
the number of times that a given tag $X$ was associated
with a specific resource as the number $f_X$ of users
who established that resource-tag association
(see also Fig.~\ref{two_resources}). That is, a broad folksonomy
has a natural notion of weight for tag associations,
which is based on social agreement.
On studying \textit{del.icio.us} we adopt a resource-centric view of
the system, that is we investigate the emergent correspondence
between a given resource and the tags that all users associate with it.
We factor out the detailed identity of the users and only deal with
the set of tags associated by the user community
with a given resource, as well as with the frequencies of occurrence
of those tags in the context of the resource.

To collect data, we used a web crawler that
connects to \textit{del.icio.us} and navigates the system's interface
as an ordinary user would do, extracting tagging metadata
and storing it for further post-processing. Our client connects
to \textit{del.icio.us} and downloads the web pages associated
with a given set of resources, using an HTML parser to extract
the tagging information from the page. The system allows to get the complete set of annotations associated with each resource.
The data used for the present analysis were retrieved in October 2006.

\section{Resource Networks from Collective Tagging Patterns}
\label{sec:distance}
In a collaborative tagging system, a set of resources defines
a ``semantic space'' that is explored and mapped by a community
of users, as they bookmark
and tag those resources~\cite{hotho2006emergent}.
We want to investigate whether the tagging activity is actually
structuring the space of resources in a semantically meaningful way,
i.e. whether partitions or subsets of resources emerge,
associated with tagging patterns that point to well-defined
meanings, areas of interest or topics. These groups of resources
could also identify, in principle, communities of users sharing
the same view of resources, or the same emergent vocabulary.

In order to gain insight into the above problem, we set up
an experiment using \textit{del.icio.us} as a data source.
We want to stress here that, since the aim of the work is to investigate whether an emergent community structure exists in folksonomy data, we are not concerned with the completeness of the dataset used. Rather, we decided to perform the experiment on the following subset: we selected two popular tags that appear to be semantically unrelated
(\textit{design} and \textit{politics}), and for each of them
we extracted from \textit{del.icio.us} a set of $200$ randomly chosen
resources (we take the first $200$ returned by the system, representing the most recently introduced by users). For each resource, we collected the complete set of annotations, i.e. all the tag assignments relative to that resource.
The corresponing dataset used for this experiment, thus consists of $400$ resources: half of them have been associated with the tag \textit{design},
while the other half has been tagged with \textit{politics}.
The idea is to construct a dataset containing at least two
semantically well-separated subsets. For each resource in the dataset,
the entire tagging history was retrieved from \textit{del.icio.us},
so that all the tag associations involving the chosen $400$
resources are known. In other words, we know how the entire
user community of \text{del.icio.us} ``categorized'' the selected
resources in terms of freely-chosen tags, with no biases
due to data collection.

\begin{figure}[htb]
\begin{center}
\includegraphics[width=0.95\columnwidth]{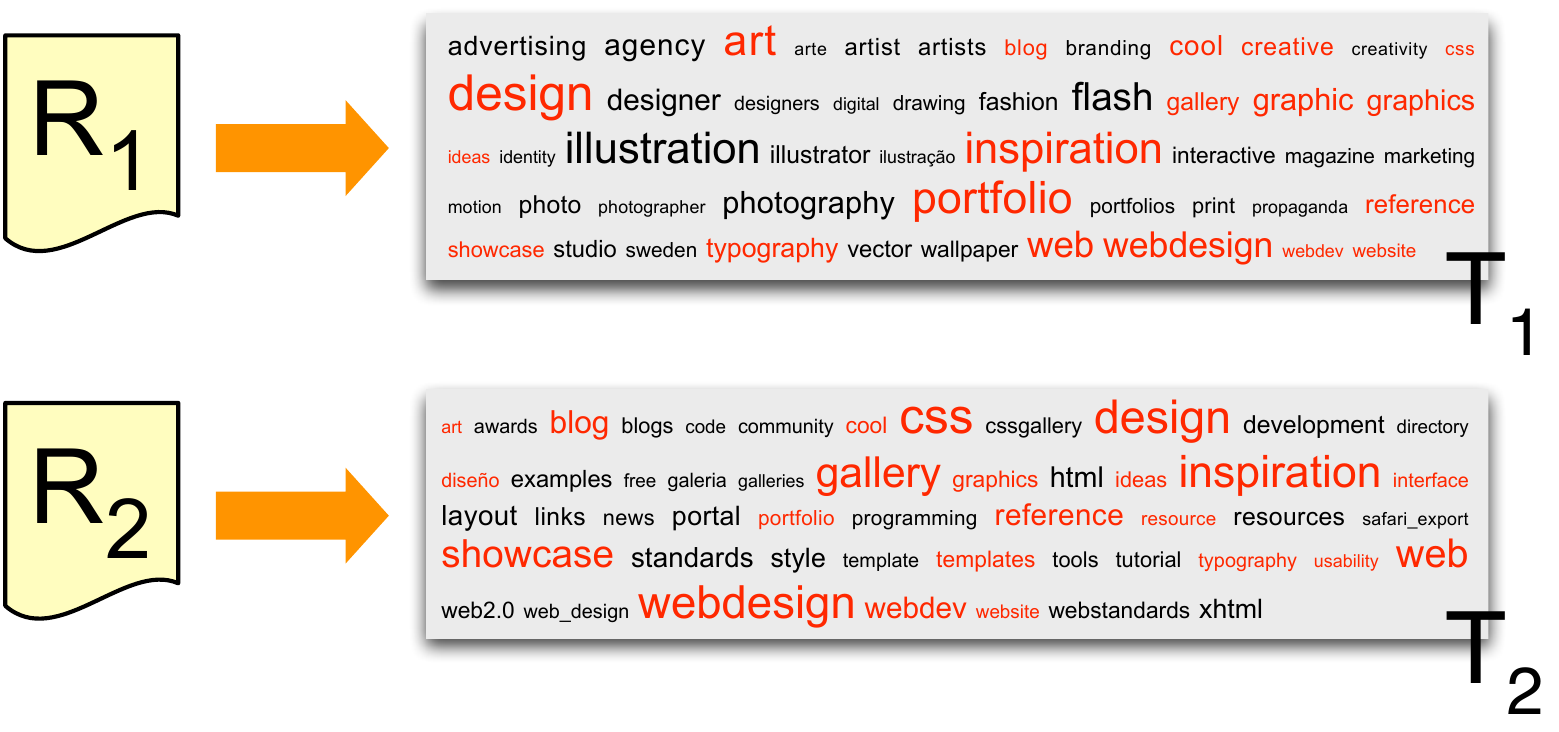}
\end{center}
\caption{The collective activity of users associates with each
resource a weighted set of tags, where the weight of a tag
is given by its frequency of occurrence in the context of a resource.
The weighted set of tags is commonly visualized by using a graphical
device called \textit{tag-cloud}: the most frequent tags
associated with a given resource are shown, and the font size of each
tag is proportional to the logarithm of its frequency of occurrence.
Our definition of similarity $w_{R_1,R_2}$ (Eq.~\ref{weight})
measures the weighted overlap between the tag-clouds associated
with the resources $R_1$ and $R_2$. Tags marked in red
belong to $T_1 \cap T_2$, the set of tags shared by the two resources.
\label{two_resources}}
\end{figure}

To uncover structures linked to specific tagging patterns
we introduce a notion of similarity between resources based
on how those resources were tagged by the user community.
For each resource, we define a \textit{tag-cloud}
as the weighted set of tags that have been used to bookmark
that resource, where the weight of tag $t$ is its frequency
of occurrence $f_t$ in the context of that resource
(Fig.~\ref{two_resources}). We want to formalize the intuitive
idea that two resources are similar if the corresponding tag-clouds
have a high degree of overlap. Given two generic resources
$R_1$ and $R_2$, and the corresponding sets of tags $T_1$
and $T_2$, a natural measure of tag-cloud overlap would be
the standard set overlap given by the cardinality of the intersection
set $T_1 \cap T_2$ divided by the cardinality of the union
set $T_1 \cup T_2$.
This simple measure, however, has a major fault: since no notion
of tag weight (frequency) is used, it is not sensitive to the social
aspects of tagging encoded in tag frequencies (and as such, it is
also vulnerable to tagging noise, 
%%%%%%%%%%%%%%%%%%%%%%%%%%%%%%%%%%%%%%%%%%%%%%%%%%%%%%%%%
%%%%%%%%%%%%%%%%%%%%%%%%%%%%%%%%%%%%%%%%%%%%%%%%%%%%%%%% 
%%%%%%%%%%%%%%%%%%%%%%%%%%%%%%%%%%%%%%%%%%%%%%%%%%%%%%%%%
i.e. errant, strange, incorrect or even malicious tagging, or
spamming~\cite{networkproperties,spamfight,heymann2007spam}).
%%%%%%%%%%%%%%%%%%%%%%%%%%%%%%%%%%%%%%%%%%%%%%%%%%%%%%%%%
%%%%%%%%%%%%%%%%%%%%%%%%%%%%%%%%%%%%%%%%%%%%%%%%%%%%%%%% 
%%%%%%%%%%%%%%%%%%%%%%%%%%%%%%%%%%%%%%%%%%%%%%%%%%%%%%%%%
To overcome this limitation
we adopt a TF-IDF weighting procedure~\cite{tfidf}. The TF-IDF weight
(Term Frequency - Inverse Document Frequency) is commonly
used in information retrieval and text mining and represents
a statistical measure used to evaluate how specific a term is
in identifying a document belonging to a collection of documents.
The importance of a term increases proportionally to the number of times
the term appears in the document, and inversely proportional to
the global frequency of the same term in the document collection.
% (red tags in Fig.~\ref{two_resources}).
%
%%% tag-clouds for two resources
We denote with $f_t^1$ and $f_t^2$
the frequencies of occurrence of tag $t$ in $T_1$ and $T_2$,
respectively, and with $f_t$ the global frequency
of tag $t$, that is the total number of times that tag $t$
was used in association with all the resources under study.

In the spirit of the TF-IDF techniques,
we normalize the frequencies of tags
by their global frequencies. When a tag is shared
by resources $R_1$ and $R_2$, it has two different
frequencies, $f_t^1$ in the context of $R_1$ and $f_t^2$
in the context $R_2$.
When performing the intersection between tag-clouds,
we use the lowest of those frequencies to define the weight
of tag $t$ in the intersection set $T_1 \cap T_2$, while we use the highest
of those frequencies when weighting the contribution of same tag
in the union set $T_1 \cup T_2$. More precisely, we define the similarity
between $R_1$ and $R_2$ as: 
\begin{equation}
w_{R_1,R_2} = \frac{\sum_{t \in T_1 \cap T_2}
\frac{\min(f_t^1, f_t^2)}{f_t}} {\sum_{t \in T_1 \cap T_2}
\frac{\max(f_t^1, f_t^2)}{f_t} + \sum_{t \in T_1-T_2 } \frac{f_t^1}{f_t} +
\sum_{t \in T_2-T_1 } \frac{f_t^2}{f_t} } \, .
\label{weight}
\end{equation}
The above expression is an extension of the simple measure
of set overlap, where the numerator is a weighted form
of set intersection and the denominator is a weighted form
of set union. By definition, $0 \leq w_{R_1,R_2} \leq 1$.
Of course the above definition is just one
of the possible similarity measures that can be employed,
and the validation of the measure we introduce here
is left to the results obtained by using it,
as shown in section~\ref{sec:communities}.
The similarity matrix introduced above can be regarded
as the adjacency matrix of a weighted network
of resources~\cite{vespignani}, where $w_{R_1,R_2}$
is the strength of the edge connecting nodes $R_1$ and $R_2$.

\begin{figure}[htb]
\begin{center}
\includegraphics[width=0.9\columnwidth]{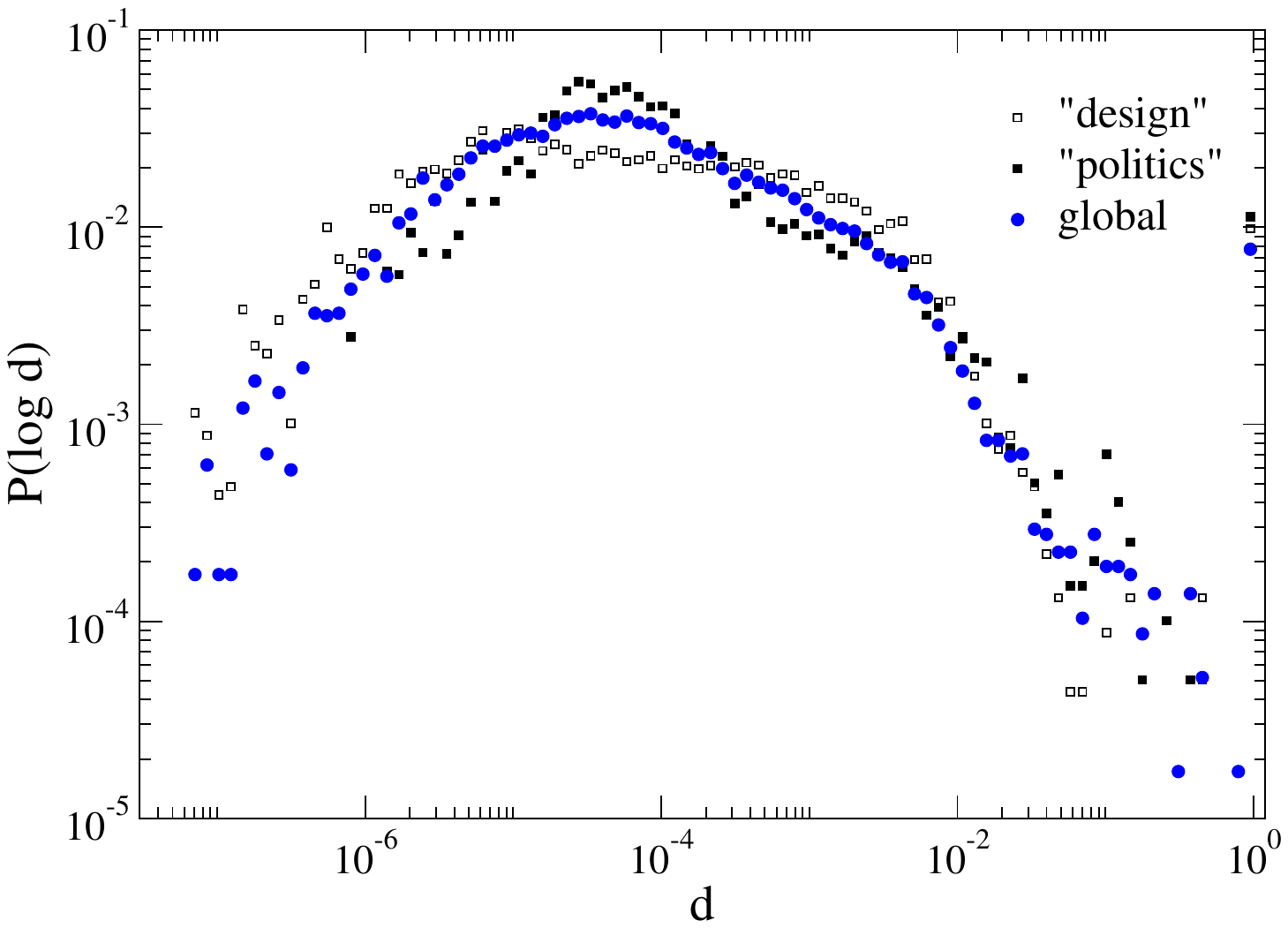}
\end{center}
\caption{Probability distributions of link strengths. The
logarithmically-binned histogram of link strengths for all pairs
of resources within a given set is displayed for three sets of
resources: empty squares correspond to resources tagged with
\textit{design}, filled squares correspond to resources tagged with
\textit{politics}, and blue circles correspond to the union of the
above sets.  It is important to observe that strength values
span several orders of magnitude, so that a non-linear
function of link strengths becomes necessary in order to capture the
full dynamic range of strength values.
\label{hysto_distance}}
\end{figure}

Fig.~\ref{hysto_distance} shows the distribution of similarities
(edge strengths in the weighted network) among all the pairs
of resources, for three different sets of resources:
the subset of resources sharing the tag \textit{design}, the subset of
resources sharing the tag \textit{politics} and the union of those sets.
Notice that the global frequency $f_t$ of a given tag $t$ depends on
the set of resources chosen for the analysis. From the plot
it is evident that weights span a wide range of values
and the logarithm of the weight is best suited
to appreciate the full range of strength variability.

%%% histogram of the weights %%%

%%% disordered matrix  %%%
\begin{figure}[htb]
\begin{center}
\includegraphics[width=\columnwidth]{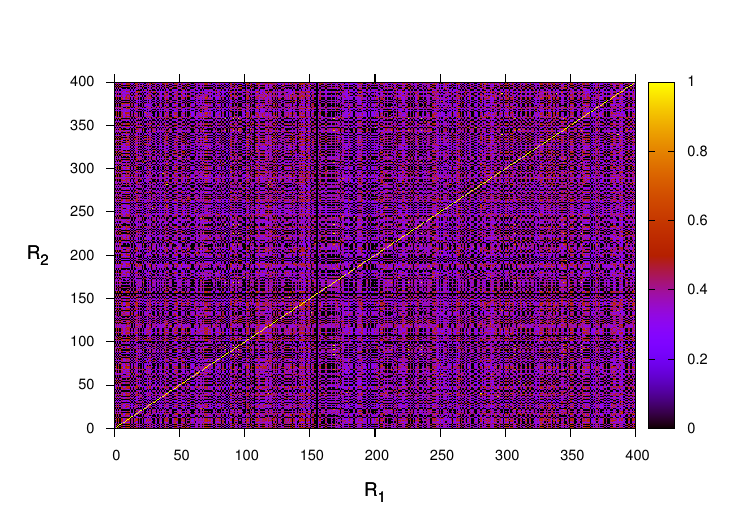}
\end{center}
\caption{Matrix $w^{\prime}$ of link strengths (Eq.~\ref{matrix_power})
for the entire set of $400$ randomly ordered
resources.  Except for the bright diagonal, whose elements are
identically equal to $1$ because of the normalization property of the
strength $w$, the matrix appears featureless. Note that no community structure
appears.
\label{disorder_matrix}}
\end{figure}

\section{Community Structure of the Resource Network}
\label{sec:communities}
In order to investigate the existence of underlying structures in the
set of resources we proceed as follows. First,
we transform the similarity matrix $w_{R_1,R_2}$ in order to compress
the dynamic range of strength values. Since the logarithmic
scale gives a good representation of the strength variability
(Fig.~\ref{hysto_distance}),
but has divergence problems in the neighborhood of zero,
we consider a matrix where each element is raised to a small (arbitrary) power
$\gamma = 0.1$.
Thus, the similarity matrix $w^{\prime}$ we will use
in the following is defined as:
% \;\;\; \forall \;\; pairs {R_1,R_2} \in U.
\begin{equation}
w^{\prime}_{R_1,R_2} = ( w_{R_1,R_2} )^\gamma \, .
\label{matrix_power}
\end{equation}

%%%%%%%%%%%%%%%%%%%%%%%%%%%%%%%%%%%%%%%%%%%%%%%%%%%%%%%%%
%%%%%%%%%%%%%%%%%%%%%%%%%%%%%%%%%%%%%%%%%%%%%%%%%%%%%%%% 
%%%%%%%%%%%%%%%%%%%%%%%%%%%%%%%%%%%%%%%%%%%%%%%%%%%%%%%%%
Note that the similarity metrics~\ref{matrix_power} is similar to
the one introduced in~\cite{filippoalex1} and~\cite{filippoalex2} for
a clastering experiment in an ontology of web pages, and was inspired by
information theory arguments.
%%%%%%%%%%%%%%%%%%%%%%%%%%%%%%%%%%%%%%%%%%%%%%%%%%%%%%%%%
%%%%%%%%%%%%%%%%%%%%%%%%%%%%%%%%%%%%%%%%%%%%%%%%%%%%%%%% 
%%%%%%%%%%%%%%%%%%%%%%%%%%%%%%%%%%%%%%%%%%%%%%%%%%%%%%%%%
 
Figure~\ref{disorder_matrix} displays the similarity matrix
(link strengths of the weighted similarity network)
between pairs of resources $w^{\prime}_{R_1,R_2}$
for the full set of $400$ resources.
The resources are randomly ordered and no structures
are visible in this representation.

The problem we have to tackle now is finding the sequence of
row and column permutations of the similarity matrix
that permits to visually identify the presence of communities of
resources, if at all possible. The goal is to obtain a matrix
with a clear visible block structure on its main diagonal.
One possible way to approach this problem is to construct
an auxiliary matrix and use information deduced from
its spectral properties to rearrange row and columns of
the original matrix. The quantity we consider is the matrix
\begin{equation}
Q = S-W \, ,
\label{matrix_laplacian}
\end{equation}
where $W_{ij}= (1 - \delta_{ij}) w_{ij}^{\prime}$
and $S$ is a diagonal matrix where each element on the main diagonal
equals the sum of the corresponding row of $W$,
i.e. $S_{ij}=\delta_{ij} \sum_j W_{ij}$.
The matrix $Q$ is non negative and resembles the Laplacian matrix of
graph theory. As shown in~\cite{capocci05,newman06}, the study of its
spectral properties can reveal the community structure of the network.

The main idea is to consider the lowest eigenvalues of $Q$. According to
the definition of $Q$, there is a always a zero eigenvalue corresponding
to an eigenvector with equal components, i.e. a trivial constant
eigenvector. Let us now consider the simple case where the matrix $Q$
is composed of exactly two non-zero blocks along its main diagonal
(i.e. with two clearly separated semantic communities). In this case,
two eigenvectors with zero eigenvalue are present, signalling
the existence of two disconnected components.
When non-zero entries connecting the two blocks are present,
only one null eigenvalue survives, and the components
of the eigenvectors with the lowest eigenvalues reveal
the community structure.
Given the set of these non trivial eigenvectors, a very simple way to
identify the communities consists in plotting their components
on a (multidimensional) scatter plot.
Each axis reports the values of the components
of the eigenvectors. In particular each point has coordinates equal
to the homologous components of one eigenvector.
In this kind of plot communities emerge as well defined
clusters of points aligned along specific directions.
The components involved in each clusters identify
the elements belonging to a given community.
Once identified the communities, it is interesting to permute the
indexes of the original matrix $W$ such that the components of the
same community become adjacent. The corresponding matrix should appear
roughly made by diagonal blocks, possibly with mixing terms
signalling an overlap between communities (blocks).

\begin{figure}[htb]
\begin{center}
\includegraphics[width=0.9\columnwidth]{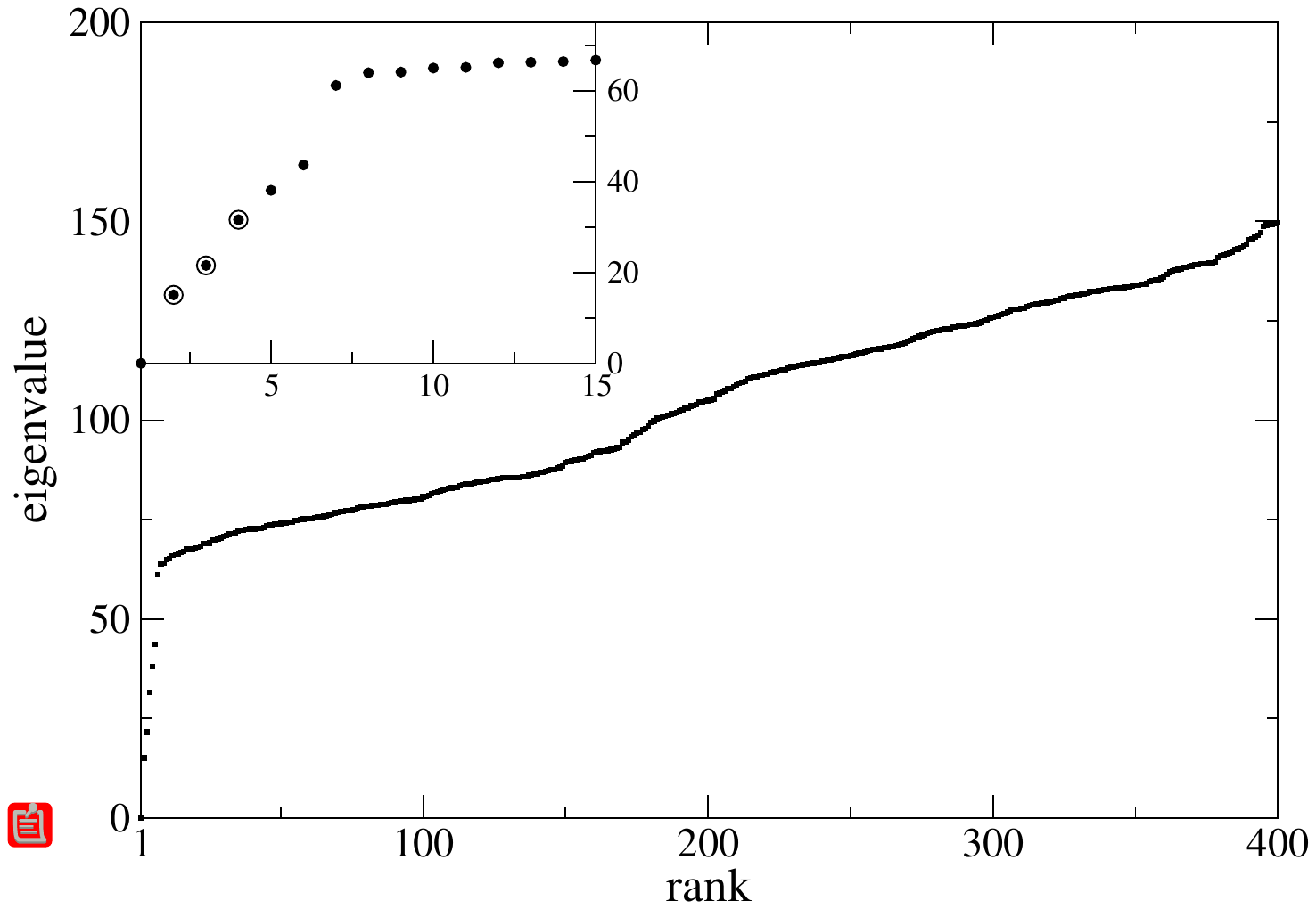}
\end{center}
\caption{Eigenvalues of the matrix $Q$ (Eq.~\ref{matrix_laplacian}).
Resource communities correspond to non-trivial eigenvalues of the spectrum,
such as the ones visible on the leftmost side of the plot and in the inset.
The three eigenvalues marked in the inset correspond to the eigenvectors 
plotted in Fig.~\ref{eigenvectors}.
\label{eigenvalues}}
\end{figure}

Figure~\ref{eigenvalues} displays the eigenvalues of $Q$
sorted by their value. As expected, the null eigenvalue is present,
corresponding to the trivial constant eigenvector.
%Higher eigenvalues correspond to non-trivial eigenvectors.

%%% eigenvalues  %%%

\begin{figure}[htb]
\begin{center}
\includegraphics[width=1.1\columnwidth]{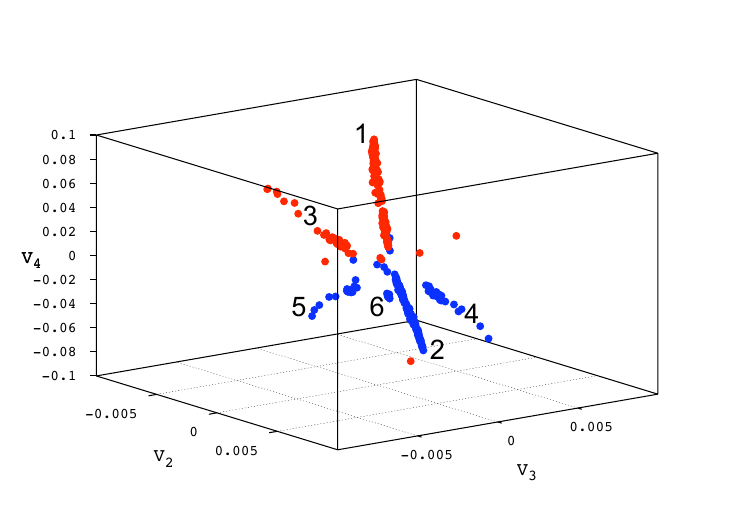}
\end{center}
\caption{Eigenvectors of the matrix $Q$ (Eq.~\ref{matrix_laplacian}).
The scatter plot displays the component values of the first
three non-trivial eigenvectors of the matrix
(marked with circles in Fig.~\ref{eigenvalues}).
The scatter plot is parametric in the component index.
Five or six clusters are visible, corresponding to the smallest non-trivial
eigenvalues of the similarity matrix.
Each cluster, marked with a numeric label, defines a community
of ``similar'' resources (in terms of tag-clouds).
Blue and red points correspond to resources tagged with
\textit{design} and \textit{politics}, respectively.
Notice that our approach clearly recovers
the two original sets of resources, and also highlights
a few finer-grained structures. Tag-clouds for the
identified communities are shown in Fig.~\ref{tagclouds}.
\label{eigenvectors}}
\end{figure}

Figure~\ref{eigenvectors} displays a 3-dimensional scatter plot
illustrating the structure of the three eigenvectors that correspond
to the three lowest non-trivial eigenvalues of $Q$ (the second, third and fourth ones, see Fig.\ref{eigenvalues}).
%%% XXXXXXXXXXXXXXXXX
The axes report the values
of the components of the second, third and fourth eigenvectors,
respectively (denoted by $V_2$, $V_3$ and $V_4$).
In particular each point has coordinates
equal to the homologous components for the three non-trivial
eigenvectors considered.
%%% XXXXXXXXXXX
The existence of at least $5$ well defined
communities is evident, with each community corresponding to
one of the five well-separated non-null eigenvalues of Fig.~\ref{eigenvalues}. A sixth very small community, corresponding to the sixth
non-trivial eigenvalue, is barely visible.

%%% eigenvectors  %%%

\begin{figure}[htb]
\begin{center}
\includegraphics[width=\columnwidth]{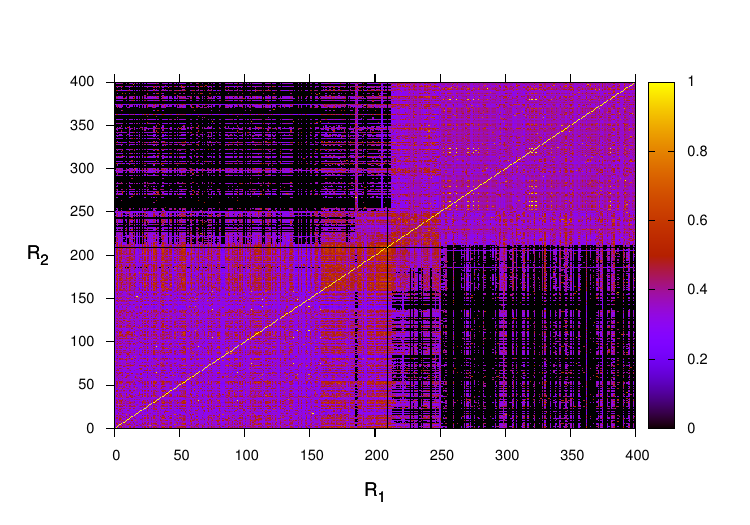}
\end{center}
\caption{Matrix $w^{\prime}$ of link strengths (see
Eq.\ref{matrix_power}) for our set of $400$ resources. Here the
resource indices are ordered by community membership (the sequence of
communities along the axes is $2$, $4$, $6$, $5$, $3$, $1$, see
Fig.~\ref{tagclouds}).  In striking contrast with
Fig.\ref{disorder_matrix}, the permutation of indices we employed
clearly exposes the community structure of the set of resources:
two large groups of resources with high-similarity, corresponding
to the blue/red rectangles at the top-right and bottom-left
of the matrix, correspond respectively to resources tagged with
\textit{design} and \textit{politics}.
On top of this, our approach also reveals the presence
of finer-grained community structures within the above
communities (red rectangular regions towards the center of the
matrix).  On direct inspection, these communities of resources
turn out to have a rather well-defined semantic characterization
in terms of tags, as shown by the tag-clouds of Fig.\ref{tagclouds}.
\label{ord_matrix}}
\end{figure}

Once we have diagonalized the matrix $Q$ the permutation of indexes
necessary to sort the component values of these eigenvectors yields
the desired ordering of rows and columns in the original matrix $W$.
By performing this reordering it is possible to visualize the matrix
of strengths of Fig.~\ref{disorder_matrix} in a way that makes it
maximally diagonal. Fig.~\ref{ord_matrix} reports the reordered matrix.

%%% ordered matrix  %%%

\begin{figure}[htb]
\begin{center}
\includegraphics[width=\columnwidth]{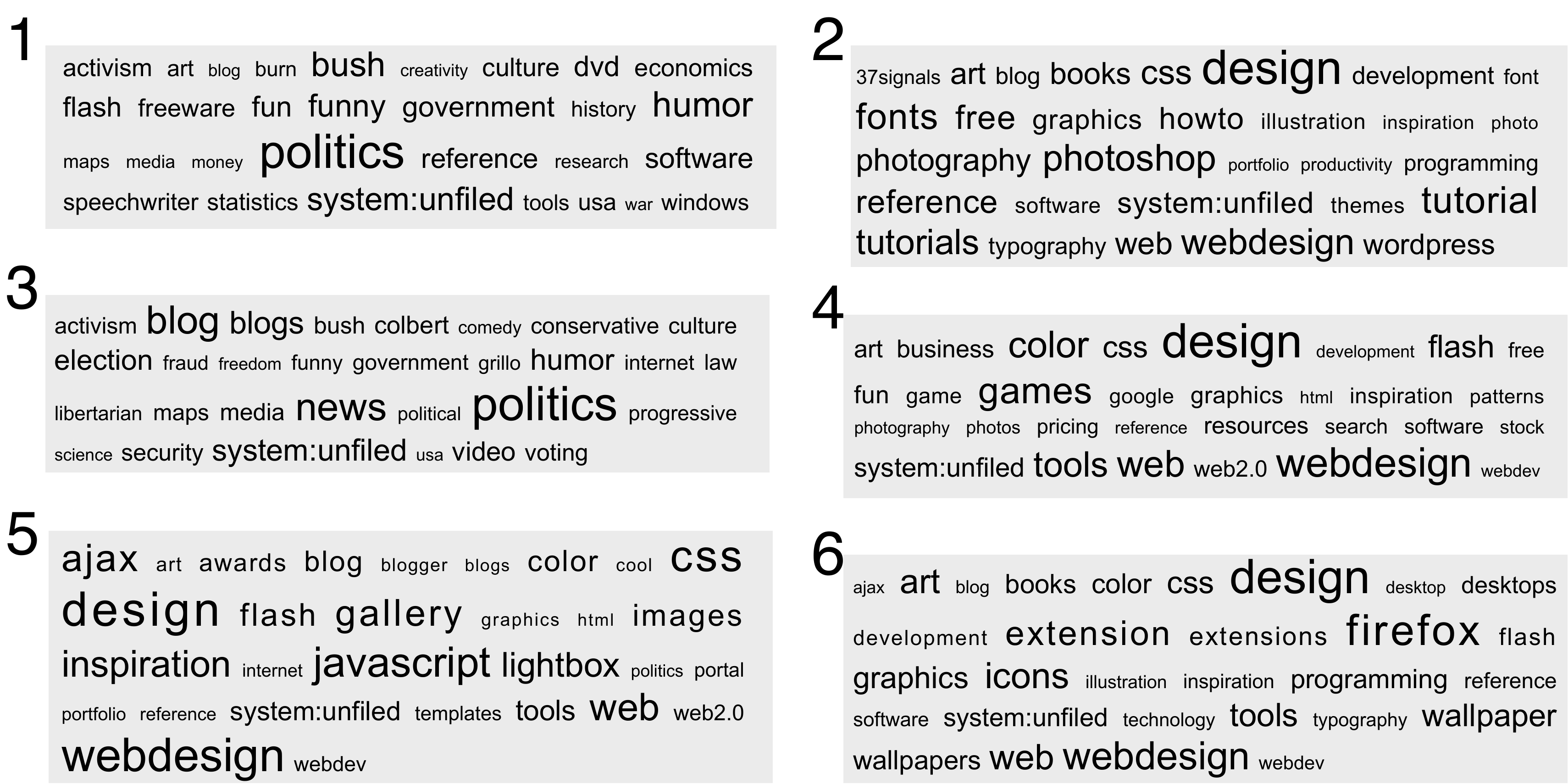}
\end{center}
\caption{Tag-clouds for the $6$ resource communities identified by our
analysis (see Fig.~\ref{eigenvectors}) ordered by decreasing
community size. Each tag-cloud shows the $30$ most frequent tags
associated with resources belonging to the corresponding community.
As usual, the size of text labels is proportional to the logarithm
of the frequency of the corresponding tag.
The first two communities (the largest ones) largely correspond
to the main division between resources tagged with \textit{politics}
and \textit{design}, respectively. Notice how each tag-cloud
is strongly characterized by only one of the above two tags.
In addition to discriminating the above two main communities,
our approach also identifies additional unexpected communities.
On inspecting the corresponding tag clouds, one can recognize
a rather well-defined semantic connotation pertaining
to each community, as discussed in the main text.
\label{tagclouds}}
\end{figure}

An interesting question is now whether the communities we have found
correspond to semantic differences in the set of resources. In order to
check this point we build for each community a tag-cloud from the tags
associated with the corresponding group of resources.
Fig.~\ref{tagclouds} reports
the six tag-clouds (ordered by decreasing number of member resources),
where the font size of each tag, as usual, is proportional
to the logarithm of its frequency of occurrence.
Despite the intrinsic difficulty of identifying the semantic context 
defined by a given tag-cloud, it is possible to recognize that each comunity
of resources -- at least for the four largest four -- comprises resources
with a specific semantic connotation.
In particular the first community can be associated
to humor in politics, the second one to visual design,
the third one to political blogs and the fourth one to web design.

%%% tag cloud globale %%%

\section{Conclusions}
\label{sec:conclusions}
The increasing impact of web-based social tools for the organization
and sharing of resources is motivating new research at the frontier
of complex systems science and computer science, with the goal
of harvesting the emergent semantics~\cite{steels} of these new tools.

The increasing interest on such new tools is based on the belief that the anarchic, uncoordinated activity of users can be used to extract meaningful and useful information. For instance, in social bookmarking systems, people  annotate personal list of resources with freely chosen tags. Wheter or not this could provide a "social" classification of resources, is the point we want to investigate with this work.
In other words, we investigate whether an emergent community structure exists in folksonomy data. To this aim,
we focused on a popular social bookmarking system
and introduced a notion of similarity between resources (annotated
objects) in terms of social patterns of tagging. We used our notion
of similarity to build weighted networks of resources,
and showed that spectral community-detection methods can be used
to expose the emergent semantics of social tagging,
identifying well-defined communities of resources that appear
associated with distinct and meaningful tagging patterns.
The present analysis was limited to an experiment where
the set of resources was artificially built by selecting
resources tagged with semantically unrelated tags: future directions
for this research include large-scale experiments on broader sets
of resources, to assess the robustness of our method,
as well as the investigation of other indicators of social agreement
that can be leveraged to expose structures in folksonomies. Such efforts could lead to improved user interfaces, increasing both usability and utility of these new powerful tools.

%\newpage

%\begin{acknowledgements}
\section*{Acknowledgements}
The authors wish to thank Melanie Aurnhammer, Andreas Hotho and Gerd
Stumme for very interesting discussions. This research has been
partly supported by the TAGora project funded by the Future and
Emerging Technologies program (IST-FET) of the European Commission
under the contract IST-34721. The information provided is the
sole responsibility of the authors and does not reflect the
Commission's opinion. The Commission is not responsible for any use
that may be made of data appearing in this publication.
%\end{acknowledgements}

%\input{referenc}

\printindex
\end{document}